\documentclass{elsart}
\usepackage{epsfig}
\usepackage{amssymb}
\begin{document}
\begin{frontmatter}
\title{\bf Compact vacuum phototriodes for operation in strong magnetic field}
\author{ M.N.Achasov\thanksref{adrr},}
\author{B.A.Baryshev,}
\author{K.I.Beloborodov,}
\author{A.V.Bozhenok,}
\author{S.V.Burdin,}
\author{V.B.Golubev,}
\author{E.E.Pyata,}
\author{S.I.Serednyakov,}
\author{Z.I.Stepanenko,}
\author{Yu.A.Tikhonov,}
\author{P.V.Vorobyov}

\thanks[adrr]{ E-mail: achasov@inp.nsk.su, FAX: +7(383-2)34-21-63}
\address{ G.I.Budker Institute of Nuclear Physics,
          Siberian Branch of the Russian Academy of Sciences,
          Novosibirsk,
          630090,
          Russia }
\date{}
\begin{abstract}
 The results of tests of 1'' vacuum phototriodes
 in a magnetic field up to 4.5~T are presented.
 It was found that output amplitude decreases by about 6\% per tesla in the
 magnetic field range from  2.0 to 4.0~T. For devices with an anode mesh pitch
 of 16$\mu$m, the output amplitude at 4.0~T  is 30\% lower than that at zero
 field.
\end{abstract}
\end{frontmatter}

\section{Introduction}
 Scintillation calorimeters which are an important part of all
 modern elementary particle detectors are often located 
 in harsh environment of strong magnetic field and high radiation fluxes.
 Photodetectors having to operate in such conditions have to satisfy special
 requirements.

 A photodetector type satisfying these requirements is the vacuum phototriode
 (VPT), which is a single-dynode photomultiplier tube with proximity focusing
 of photoelectrons. This makes it possible to operate in a strong axial
 magnetic field. Use of radiation-hard glass for VPT manufacturing makes it
 also tolerant to high doses of ionizing radiation. VPTs are already widely
 used in calorimetry, for example in the detectors DELPHI \cite{delph1,delph2}
 and OPAL \cite{opal} at LEP. They are also planned for use in the end-cap PbWO
 calorimeter of the CMS detector for LHC \cite{cms}.

 Another type of photodetector for calorimetry, the avalanche photodiode (APD),
 shows also excellent performance in strong magnetic field \cite{apd1}. APDs
 have however some drawbacks, such as a small active area and direct counting
 of charged particles which can make VPTs preferable in some cases.

 Several years ago VPTs with the capability to work in magnetic fields up to
 2.5~T \cite{vpt1} were developed in BINP; about 3500 devices of this kind
 operate in the electromagnetic calorimeters of the SND \cite{snd}, CMD-2
 \cite{cmd} and KEDR \cite{kedr} detectors. In 1999 a new low cost VPT was
 developed for operation in strong magnetic field. In this paper the results
 of tests of the new devices in a magnetic fields up to 4.5~T are presented,
 and its performance compared with that of similar photodetectors described in
 \cite{tetrod,lenin}. 

\section{Phototriode parameters and experimental set-up}

 The VPTs were manufactured using conventional bulb technology described
 elsewhere in \cite{vpt1}. All VPT electrodes are connected to pins on its
 base. The semitransparent bialkali photocathode was formed on the inner
 surface of a window glass \mbox{S52-1} or \mbox{S52-2} with transparency of
 more than 96\% in the wavelength range from 300 to 900 nm. The quantum
 efficiency of the photocathode in the maximum of spectral sensitivity
 measured with calibrated light source was about 20\%.

 The VPT performance in magnetic field depends on the anode mesh
 pitch. The smaller the pitch, the larger fraction of secondary emission
 electrons from the dynode is collected on the anode, but on the other hand a
 smaller part of accelerated photoelectrons reach the dynode. Prototypes with
 the anode mesh pitch $s$ of 250, 100, 50 and 16 $\mu$m were manufactured. The
 devices height is 40 mm, tube diameter is 25 mm, the photocathode spectral
 sensitive region is from 360 to 600 nm, maximum of photocathode spectral
 sensitivity is $\lambda_{max} = 420$ nm, total photocathode sensitivity is
 95$\mu$A/lm, typical quantum efficiency at $\lambda_{max}$ is about 20\%,
 dark current is less than 1 nA, gain without magnetic field is 15 for
 $s=50\mu$m and 10 for $s=16$ $\mu$m, anode mesh transparency is 60\%  for
 $s=50$ $\mu$m and 52\% for $s=16$ $\mu$m.

\begin{figure}[h]
\begin{center}
\epsfig{figure=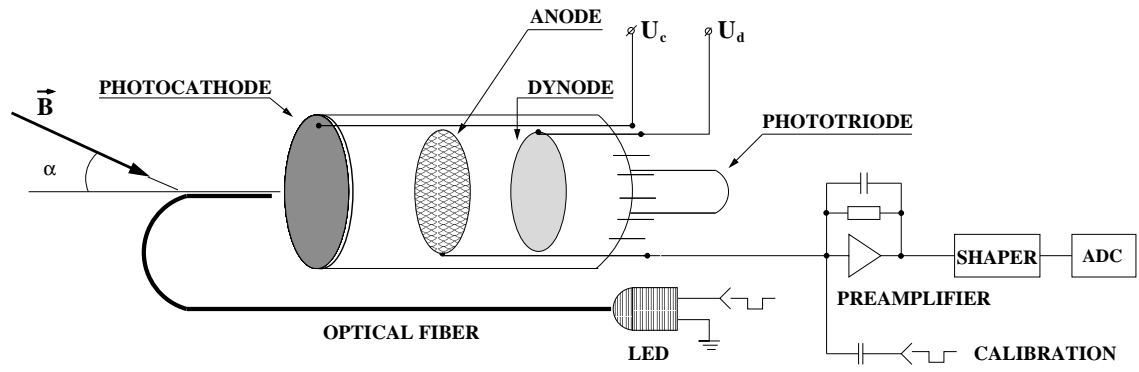,width=15cm}
\caption{Test system layout}
\label{scheme}
\end{center}
\end{figure}

 The layout of the test system used to check the operation in high magnetic 
 fields is shown in Fig.\ref{scheme}. Measurements were performed using a
 charge sensitive preamplifier with a sensitivity of 0.7~V/pC and a shaper
 with integration and differentiation time constants of 2$\mu$s. Green LED with
 wavelength of about 520 nm was used as a source of light signals. A magnetic
 field with a strength up to $4.5\mbox{T}$ was produced by a superconducting
 solenoid. The VPT axis could be tilted by up to 30 degrees with respect to
 the magnetic field direction.

\begin{figure}[h]
\begin{center}
\epsfig{figure=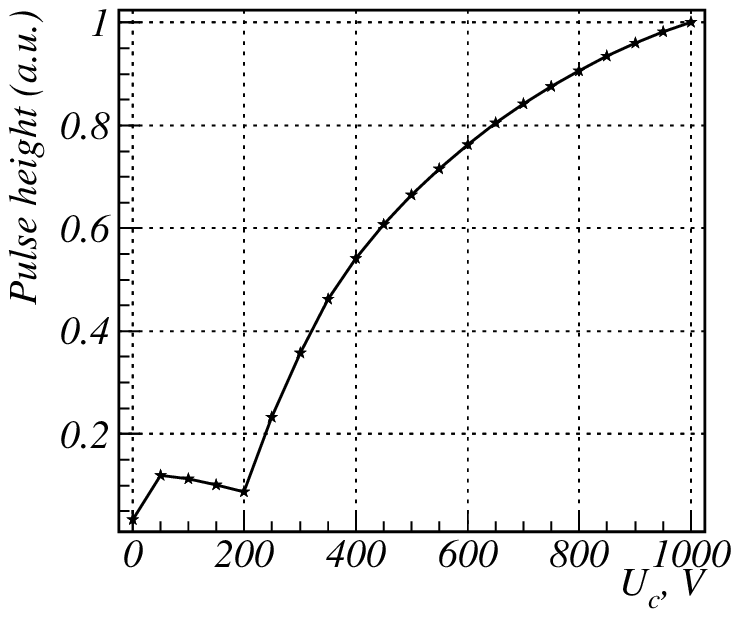,height=7cm}
\caption{The VPT output signal as a function of photocathode voltage $U_c$.
The dynode voltage is fixed to $U_d = -200\mbox{V}$. The kink at $U_c=-200$~V
reflects the transition from photodiode to a phototriode operation mode
of the device.}
\label{uc}
\vspace{3mm}
\epsfig{figure=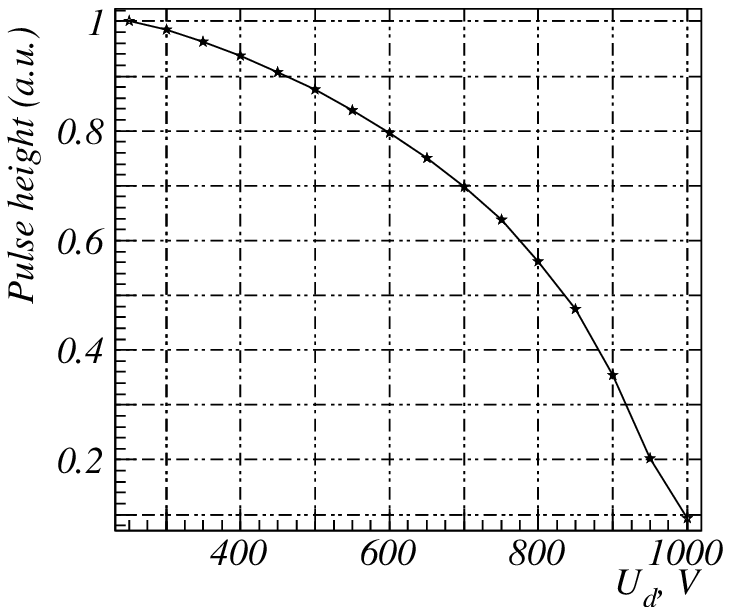,height=7cm}
\caption{The VPT output signal as a function of dynode voltage $U_d$.
         The photocathode voltage is fixed to $U_c = -1000\mbox{V}$}
\label{ud}         
\end{center}         
\end{figure}

 The dependences of output signal on photocathode and dynode voltages at zero
 magnetic field are shown in Figs.\ref{uc},\ref{ud}. For further measurements
 the photocathode and dynode voltages were fixed to $U_c=-1000\mbox{V}$ and
 $ U_d=-200\mbox{V}$ respectively.

\begin{figure}[h]
\begin{center}
\epsfig{figure=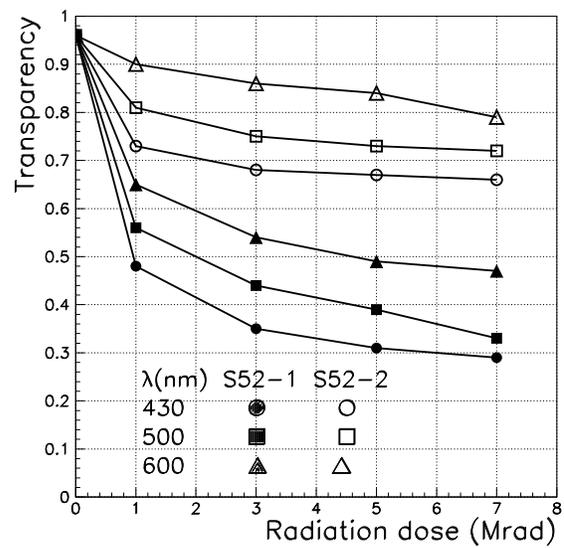,width=8cm}
\caption{Transparency of S52-1 and S52-2 glasses, 1 mm thick,
         as a function of absorbed radiation dose for different wavelength.}
\label{radst}         
\end{center}         
\end{figure}

 After absorption of high dose of radiation the input glass window may darken
 thus decreasing the VPT sensitivity. The dependence of transparency
 of VPT windows made of S52-1 and S52-2 glass on the radiation dose is shown in
 Fig.\ref{radst}. The radiation harder S52-2 glass was chosen
 as a material for the VPT window.
 
\section{VPT performance in magnetic field}

\begin{figure}[h]
\begin{center}
\epsfig{figure=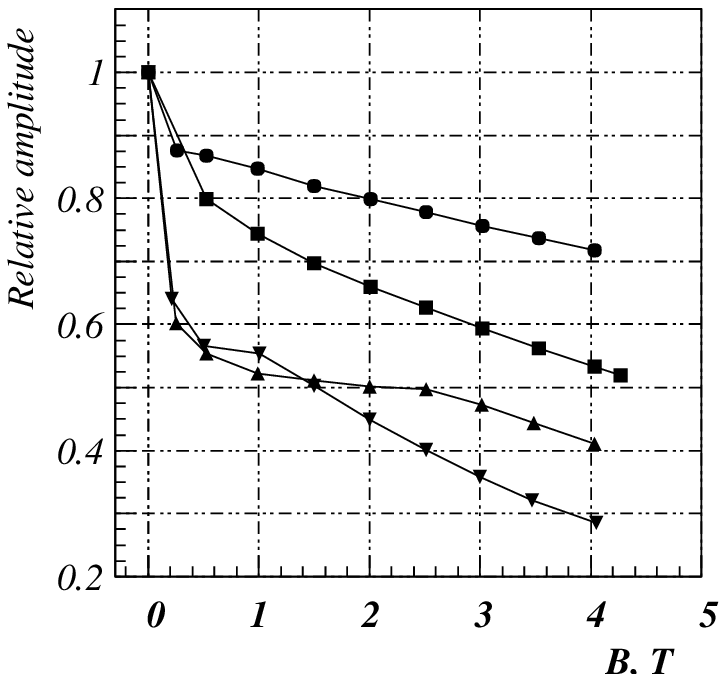,height=7cm}
\caption{Relative output amplitude as a function of magnetic field at
         $\alpha=0^\circ$ for VPTs with anode mesh spacing 16$\mu$m
	 ($\bullet$), 50$\mu$m ($\blacksquare$), 100$\mu$m ($\blacktriangle$),
         250$\mu$m ($\blacktriangledown$). Full photocathode illumination.}
\label{nda0b}
\epsfig{figure=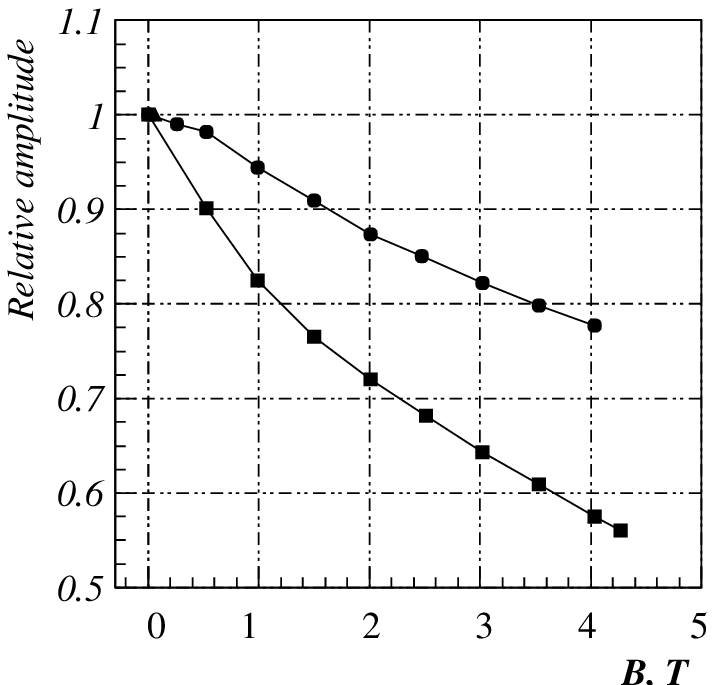,height=7cm}
\caption{Relative output amplitude as a function of magnetic field at
         $\alpha=0^\circ$ for VPTs with anode mesh spacing 16$\mu$m
	 ($\bullet$), 50$\mu$m ($\blacksquare$). Illumination of the central
	 part of photocathode.}
\label{eda0b}         
\end{center}         
\end{figure}

 The dependence of VPT output amplitude on the magnetic
 field strength was measured both illuminating the entire photocathode
 area, and its central part  (10mm in diameter). Fig.\ref{nda0b} shows the
 dependences of the output signal on magnetic field for VPTs with different
 anode mesh pitches in case of illumination of entire photocathode.
 The amplitude drop at $B=4.0\mbox{T}$ varies from 70\%  for tubes with 
 $s=250\mu$m mesh to 30\% for $s=16\mu$m. The dependence for $s=50\mu$m and
 $s=16\mu$m with illumination of the central part of photocathode is shown in
 Fig.\ref{eda0b}. The output signal decreases by about 6\% per tesla in a range
 of field from 2.0 to 4.0T. The difference in amplitude drops for illumination
 of the full photocathode area and of its central part can be explained
 by effective cut-off of the peripheral area of the photocathode in
 axial magnetic fields. Photoelectrons from this area, propagating
 along the magnetic field, cannot reach the dynode which due to
 manufacture constraints has smaller diameter.
 
\begin{figure}[h]
\begin{center}
\epsfig{figure=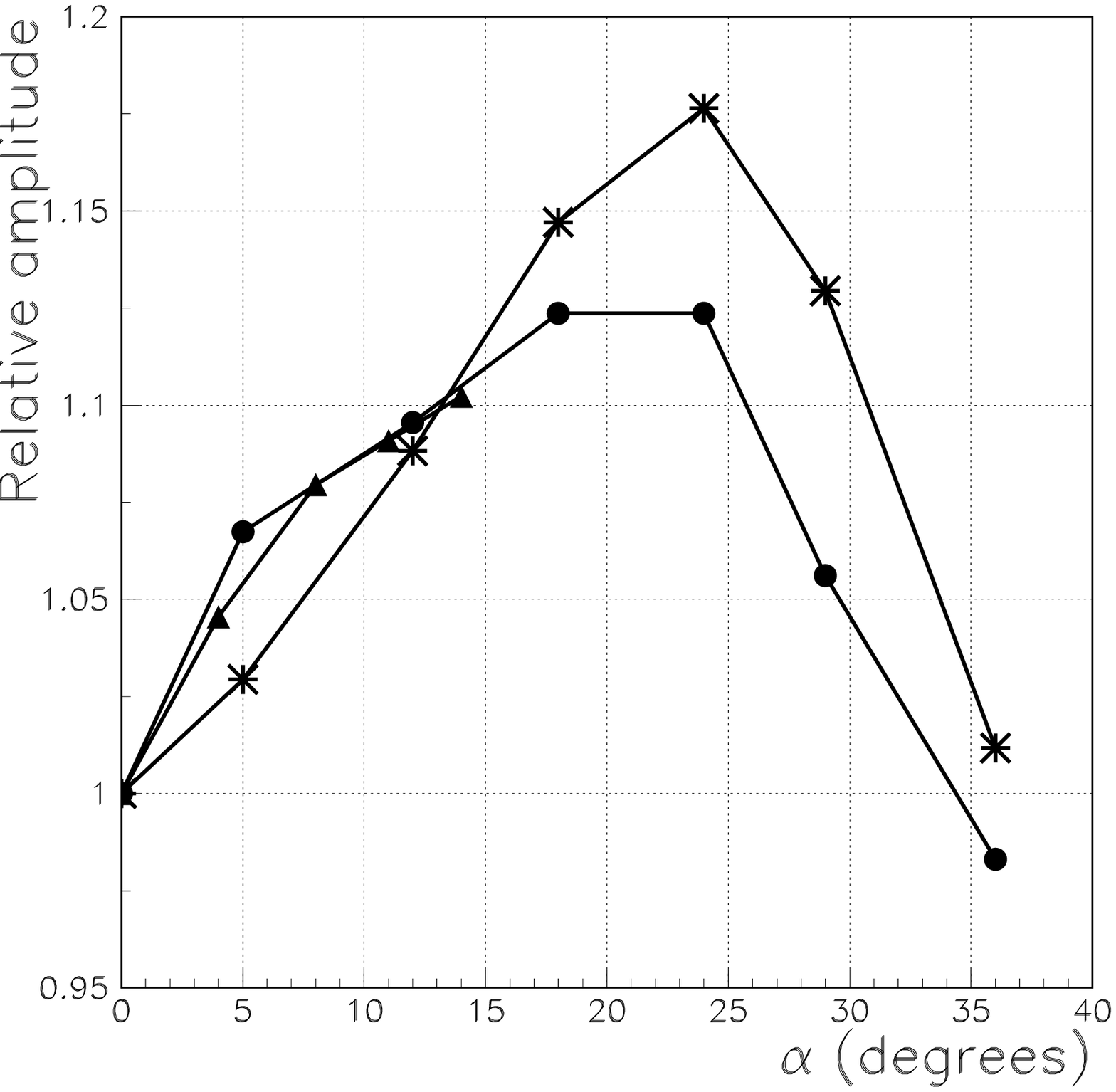,height=8cm}
\caption{Relative output amplitude as a function of
         tilt angle $\alpha$ in 4T
         magnetic field for VPTs with anode mesh spacing 16$\mu$m ($\bullet$),
	 100$\mu$m ($\blacktriangle$) in case of full photocathode
	 illumination. The line through ($\ast$) corresponds to VPT with mesh
	 spacing 16$\mu$m, when only the central part of photocathode was
	 illuminated.}
\label{vang}
\epsfig{figure=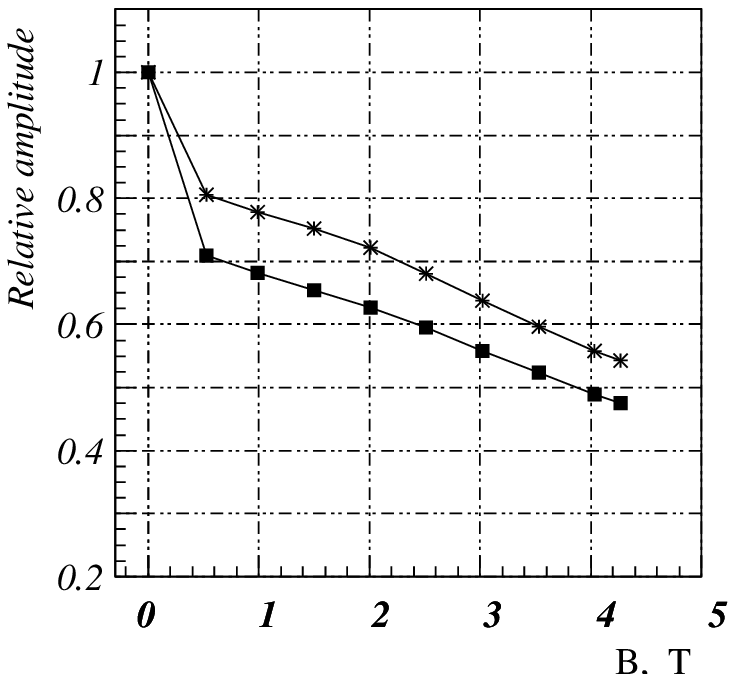,height=7cm}
\caption{Relative output amplitude as a function of magnetic field at
         $\alpha=30^\circ$ for VPT with anode mesh spacing 50$\mu$m and full
	 photocathode illumination  ($\blacksquare$) or illumination of the
	 central part only ($\ast$).}
\label{tda3b}                  
\end{center}                  
\end{figure}

 The dependence of the output amplitude on $\alpha$, the angle between the
 magnetic field and the tube axis, is shown in Fig.\ref{vang}.
 The initial amplitude increase by $\sim$ 15\%
 with $\alpha$ can be attributed to the increase of the secondary electron
 emission coefficient on the dynode for larger impact angles of the 
 photoelectrons. At larger tilt angles another effect, the decrease of anode
 mesh transparency for photoelectrons, apparently becomes dominant.
 The amplitude dependence for mesh pitch $s=50\mu$m and $\alpha=30^\circ$ on
 the magnetic field strength is shown in Fig.\ref{tda3b}.

 The tests demonstrate that the VPTs with $16\mu$m anode mesh are the best for
 operation in strong magnetic field. The output signal decreases by less than
 30\%  at 4.0~T, for a angles between the tube axis and the field up to
 $40^\circ$. Recently, the results of the RIE VPTs (diameter of 25 mm) tests
 in a magnetic field were presented in Ref.\cite{lenin}. The output amplitude
 of the device making use of a fine mesh with 100 lines per mm decreases by
 about 40\% at 4T and $\alpha=0^\circ$. As reported in Ref.\cite{tetrod},
 in commercial Hamamatsu 25 mm vacuum phototetrodes the output signal amplitude
 decrease by about 70\% in the same conditions.
  
\section{Conclusions}

 We describe the development of prototypes of compact vacuum phototriodes with
 quantum efficiency of  $\sim$ 20\%  and gain $10\div15$ for operation in
 strong magnetic field. Their performance in the fields up to 4.5~T was tested.
 It was found that the decrease of output amplitude is about 6\% per tesla in
 the magnetic field range from  2.0 to 4.0~T. For VPT with anode mesh pitch of
 16$\mu$m the output amplitude at 4.0~T is 30\% less than that without magnetic field.

\newpage

\end{document}